\documentclass[a4paper]{article}
\usepackage{epsfig,amsmath,amsfonts,amssymb,setspace,multirow,textcomp}
\usepackage[T1]{fontenc}
\makeatletter
\renewcommand{\@oddhead}{\textit{Advances in Astronomy and Space Physics} \hfil}
\renewcommand{\@evenfoot}{\hfil \thepage \hfil}
\renewcommand{\@oddfoot}{\hfil \thepage \hfil}
\makeatother

\renewenvironment{thebibliography}[1]{\begin{oldthebibliography}{#1}\setlength{\parskip}{0ex}\setlength{\itemsep}{0ex}}{\end{oldthebibliography}}

\begin{document}
\fontsize{11}{11}\selectfont 
\title{Nonthermal emission model of isolated X-ray pulsar RX J0420.0-5022}
\author{\textsl{N.~Chkheidze$^{1}$, Iu.~Babyk$^{2,3,4}$}}
\date{\vspace*{-6ex}}
\maketitle
\begin{center} {\small $^{1}$Center for theoretical Astrophysics, ITP, Ilia State
University, 0162, Tbilisi, Georgia\\
$^{2}$Main Astronomical Observatory of NAS of Ukraine, Zabolotnogo str., 27, 03680, Kyiv, Ukraine\\
$^{3}$Dublin Institute for Advanced Studies, Fitzwilliam Place., 31, Dublin 2, Ireland\\
$^{4}$Dublin City University, Dublin 9, Ireland\\
{\tt nino.chkheidze@iliauni.edu.ge, babikyura@gmail.com}}
\end{center}

\begin{abstract}
In the present paper an alternative theoretical interpretation to
the generally assumed thermal emission models of the observed X-ray
spectrum of isolated pulsar RX J0420.0-5022 is presented. It is well
known that the distribution function of relativistic particles is
one-dimensional at the pulsar surface. However, cyclotron
instability causes an appearance of transverse momenta of
relativistic electrons, which as a result, start to radiate in the
synchrotron regime. On the basis of the Vlasov's kinetic equation we
study the process of the quasi-linear diffusion (QLD) developed by
means of the cyclotron instability. This mechanism provides
generation of optical and X-ray emission on the light cylinder
lengthscales. The analysis of the three archival XMM-Newton
observations of RX J0420.0-5022 is performed. Considering a
different approach of the synchrotron emission theory, the spectral
energy distribution is obtained that is in a good agreement with the
observational data. A fit to the X-ray spectrum is performed using
both the present synchrotron emission model spectrum absorbed by
cold interstellar matter and generally assumed absorbed black-body
model.
\\[1ex]

\end{abstract}

\section*{\sc introduction}
\indent \indent The soft X-ray source {\it RXJ0420.0-5022} (hereafter RXJ0420) belongs to the group of seven radio-quiet, isolated neutron stars discovered in the ROSAT all-sky survey data and often called the X-ray Dim Isolated Neutron Stars (XDINSs, see \cite{hab07} for review). In the optical domain, only a few XDINSs have counterparts certified by their proper motion measurements and likely candidates are proposed for remaining, based on their coincidence with the X-ray positions \cite{mig}. The X-ray spectra of XDINSs are well represented by an absorbed black-body ($kT\approx45-100$eV) which is emitted by the hotter part of the neutron star surface \cite{hab07}. Whereas, the cooler part of the XDINSs surface is assumed to be emitting in the optical domain \cite{po}. The formation of a non-uniform distribution of the surface temperature is still under debate, which is more likely artificial and needs to be examined by convincing theory. The optical emission is also explained in terms of hydrogen layers of finely tuned thickness superposed on a condensed matter surface reprocessing the surface radiation \cite{ho07}, or of non-thermal emission from particles in the star magnetosphere \cite{mo03}.

Alternatively to somewhat contrived thermal emission models the
observed properties of the faintest object in X-rays among the
XDINSs RXJ0420 can be explained in the framework of the plasma
emission model. This model has already been applied to explain the
observed spectra of two other XDINSs (RX J1856.5-3754 and RBS1774,
see \cite{ch11,new} for more details). The plasma emission model is
based on the well-developed theory of pulsars \cite{lomi,mach79} and
suggests successful interpretation of the observational data.
According to these works, in the electron-positron plasma of a
pulsar magnetosphere the low frequency cyclotron modes, on the
quasi-linear evolution stage create conditions for generation of the
high energy synchrotron emission. Generally speaking, at the pulsar
surface relativistic particles efficiently loose their perpendicular
momenta via synchrotron emission in very strong magnetic fields and
therefore, they very rapidly transit to their ground Landau state
(pitch angles are vanishing). However, the cyclotron instability
causes appearance of transverse momenta of relativistic particles in
the outer parts of the pulsar magnetosphere. Therefore, the resonant
electrons start to radiate in the synchrotron regime.

We suppose that the observed X-ray emission of RXJ0420 is generated by the synchrotron mechanism at the light cylinder length-scales. In general the synchrotron radiation spectrum is considered to be a power-law, which is not consistent with the observational data from RXJ0420. The standard theory of the synchrotron radiation \cite{bekefi,ginz81} typically provides the power-law spectral
energy distribution. Contrary to the standard scenario, by taking into account the specifications produced due to the present emission model we obtain different spectral distribution, which can be successfully fitted with the measured X-ray spectrum of RXJ0420.

In the present paper, we describe the plasma emission model and derive the synchrotron radiation spectrum based on our scenario
(Sec. 2), present the results of spectral analysis of re-processed XMM-Newton archival data from RXJ0420 (Sec.3), and make conclusions (Sec. 4).

\section*{\sc emission model}

\indent \indent Any well known theory of pulsar emission suggests that, the observed
radiation is generated due to processes taking place in an
electron-positron plasma. It is generally assumed that the pulsar
magnetosphere is filled by dense relativistic electron-positron
plasma with  an anisotropic one-dimensional distribution function
(see Fig. 1 from \cite{arons}) and consists of the following
components: a bulk of plasma with an average Lorentz-factor
$\gamma_{p}\simeq10^{2}$, a tail on the distribution function with
$\gamma_{t}\simeq10^{5}$, and the primary beam with
$\gamma_{b}\simeq10^{7}$. The distribution function is
one-dimensional and anisotropic and plasma becomes unstable, which
might cause excitation of the plasma eigen modes. While considering
the eigen-modes of the electron-positron plasma having small
inclination angles with respect to the magnetic field, one has three
branches, two of which are mixed longitudinal-transversal waves
($lt_{1,2}$). The high frequency branch on the diagram
$\omega(k_{_\parallel})$ begins with the Langmuir frequency and for
longitudinal waves ($k_{\perp}=0$), $lt_1$ reduces to the pure
longitudinal Langmuir mode. The low frequency branch, $lt_2$, is
similar to the Alfv\'en wave. The third $t$ mode, is the pure
transversal wave, the electric component of which ${\bf E^t}$ is
perpendicular to the plane  of the wave vector, and the magnetic
field, $({\bf k,B_0})$. The vector of the electric field, ${\bf
E^{lt_1,lt_2}}$ is located in the plane $({\bf k,B_0})$. When
$k_{\perp}=0$, the $t$-mode merges with the $lt$ waves and the
corresponding spectra is given by \cite{kmm}
\begin{equation}\label{1}
\omega_t \approx kc\left(1-\delta\right),\;\;\;\;\; \delta =
\frac{\omega_p^2}{4\omega_B^2\gamma_p^3}
\end{equation}
where $k$ is the modulus of the wave vector, $c$ is the speed of
light, $\omega_p \equiv \sqrt{4\pi n_pe^2/m}$ is the plasma
frequency, $e$ and $m$ are the electron's charge and the rest mass,
respectively, $n_p$ is the plasma density, $\omega_B\equiv eB/mc$ is
the cyclotron frequency and $B$ is the magnetic field induction.

The main mechanism of wave generation in plasmas of the pulsar
magnetosphere is the cyclotron instability. The cyclotron resonance
condition can be written as \cite{kmm}:
\begin{equation}\label{1}
    \omega-k_{\|}V_{\|}-k_{x}u_{x}+\frac{\omega_{B}}{\gamma_{r}}=0,
\end{equation}
where $u_{x}=cV_{\varphi}\gamma_{r}/\rho\omega_{B}$ is the drift
velocity of the particles due to curvature of the field lines with
the curvature radius $\rho$. During the wave generation process, one
also has a simultaneous feedback of these waves on the resonant
electrons \cite{vvs}. This mechanism is described by the QLD,
leading to a diffusion of particles as along as across the magnetic
field lines. Therefore, resonant particles acquire transverse
momenta (pitch angles) and, as a result, start to radiate through
the synchrotron mechanism.

The wave excitation leads to redistribution process of the resonant
particles via the QLD. The kinetic equation for the distribution
function of the resonant electrons can be written as \cite{ch11}:
\begin{eqnarray}\label{2}
\frac{\partial\textit{f }^{0}}{\partial
    t}+\frac{\partial}{\partial
p_{\parallel}}\left\{F_{\parallel}\textit{f
}^{0}\right\}+\frac{1}{p_{\perp}}\frac{\partial}{\partial
p_{\perp}}\left\{p_{\perp}F_{\perp}\textit{f }^{0}\right\}=\nonumber
\\=\frac{1}{p_{\perp}}\frac{\partial}{\partial p_{\perp}}\left\{p_{\perp}D_{\perp,\perp}\frac{\partial}{\partial p_{\perp}}\textit{f }^{0}\left(\mathbf{p}\right)\right\}.
\end{eqnarray}
where
\begin{equation}\label{4}
    F_{\perp}=-\alpha_{s}\frac{p_{\perp}}{p_{\parallel}}\left(1+\frac{p_{\perp}^{2}}{m^{2}c^{2}}\right),\qquad
    F_{\parallel}=-\frac{\alpha_{s}}{m^{2}c^{2}}p_{\perp}^{2},
\end{equation}
are the transversal and longitudinal components of the synchrotron
radiation reaction force and
$\alpha_{s}=2e^{2}\omega_{B}^{2}/3c^{2}$.

Here $D_{\perp,\perp}$ is the perpendicular diffusion coefficient,
which can be defined as follows \cite{ch11}
\begin{eqnarray}\label{16}
    D_{\perp,\perp}=\frac{e^{2}}{8c}\delta|E_{k}|^{2},
\end{eqnarray}
where $|E_{k}|^{2}$ is  a density of electric energy in the waves
and its value can be estimated from the expression $
|E_{k}|^{2}\approx mc^{2}n_{b}\gamma_{b}c/2\omega$, where $\omega$
is the frequency of the cyclotron waves.

The transversal QLD increases the pitch-angle, whereas force
$F_{\perp}$ resists this process, leading to a stationary state
($\partial\textit{f}/\partial t=0$). The pitch-angles acquired by
resonant electrons during the process of the QLD satisfies $\psi=
p_{\perp}/p_{\parallel}\ll1$. Thus, one can assume that
$\partial/\partial p_{\perp}>>\partial/\partial p_{\parallel}$. In
this case the solution of Eq.(3) gives the distribution function of
the resonant particles by their perpendicular momenta \cite{ch11}

\begin{equation}\label{9}
    \textit{f}(p_{\perp})=C exp\left(\int
    \frac{F_{\perp}}{D_{\perp,\perp}}dp_{\perp}\right)=Ce^{-\left(\frac{p_{\perp}}{p_{\perp_{0}}}\right)^{4}},
\end{equation}
where
\begin{equation}\label{11}
     p_{\perp_{0}}\approx\frac{\pi^{1/2}}{B\gamma_{p}^{2}}\left(\frac{3m^{9}c^{11}\gamma_{b}^{5}}{32e^{6}P^{3}}\right)^{1/4}.
\end{equation}
And for the mean value of the pitch angle we find $\psi_0\approx
p_{\perp_{0}}/p_{\parallel}\simeq10^{-3}$. Synchrotron emission is
generated as the result of appearance of pitch angles.

\subsection*{\sc synchrotron radiation spectrum}

To explain the observed X-ray emission of RXJ0420, let us assume
that the resonant particles are the primary beam electrons with
$\gamma_{b}\sim10^{7}$, giving the synchrotron radiation in the soft
X-ray domain. According to our emission scenario, the synchrotron
radiation is generated as the result of acquirement of pitch angles
by resonant particles, during the QLD stage of the cyclotron
instability. As was shown in \cite{malov02}, the cyclotron
resonance condition (see Eq. (2)) is fulfilled at the light cylinder
lengthscales. Consequently, the observed X-ray emission comes from
the region near the light cylinder, where the geometry of the field
lines is determined by the curvature drift instability excited at
the same lengthscales \cite{os09}. The curvature drift instability
effectively rectifies the magnetic field lines (the curvature radius
tends to infinity). Therefore, in the synchrotron emission
generation region the field lines are practically straight and
parallel to each other and one can assume that electrons with
$\psi\approx\psi_0$ efficiently emit in the observer's direction.

The synchrotron emission flux of the set of electrons in this case
is given by (see \cite{ch11})
\begin{equation}\label{14}
    F_{\epsilon}\propto\int\textit{f}_{\parallel}(p_{\parallel})B\psi_{0}\frac{\epsilon}{\epsilon_{m}}\left[\int_{\epsilon/
    \epsilon_{m}}^{\infty}K_{5/3}(z)dz \right] dp_{\parallel}.
\end{equation}
Here $\textit{f}_{\parallel}(p_{\parallel})$ is the distribution
function of the resonant electrons by their parallel momenta,
$\epsilon_{m}\approx5\cdot10^{-12}B\psi\gamma^{2}$keV is a photon
energy of maximum of the synchrotron spectrum of a single electron
and $K_{5/3}(z)$ is the Macdonald function.

In order to find the synchrotron emission spectrum, one needs to
know the behaviour of the distribution function,
$\textit{f}_{_{\parallel}}(p_{_{\parallel}})$. For solving this
problem, we consider the equation governing the evolution of
$\textit{f}_{_{\parallel}}(p_{_{\parallel}})$ \cite{ch11}
\begin{equation}\label{fpar}
    \frac{\partial\textit{f}_{_{\parallel}}}{\partial t}=\frac{\partial}{\partial
    p_{_{\parallel}}}\left({\frac{\alpha_{s}}{m^{2}c^{2}\pi^{1/2}}p_{\perp_{0}}^{2}\textit{f}_{_{\parallel}}}\right).
\end{equation}
Considering the quasi-stationary case from Eq. (9) we can find the
redistributed distribution function of the resonant particles by
their parallel momenta
\begin{eqnarray}\label{19}
    \textit{f}_{\parallel}\propto\frac{1}{p_{\parallel}^{1/2}|E_{k}|}.
\end{eqnarray}

On the other hand, the cyclotron noise is described by the equation
\begin{equation}\label{20}
    \frac{\partial|E_{k}|^{2}}{\partial
    t}=2\Gamma_{c}|E_{k}|^{2},
\end{equation}
where
\begin{equation}\label{21}
   \Gamma_{c}=\frac{\pi^{2}e^{2}}{k_{\parallel}}\textit{f}_{\parallel}(p_{r}),
\end{equation}
is the growth rate of the instability and from the resonance
condition (2) it follows that
$k_{\parallel}\approx\omega_{B}/c\delta\gamma_{r}$.

Combining Eqs. (9) and (11) it is easy to find that \cite{ch11}:
\begin{equation}\label{27}
     |E_{k}|^{2}\propto p_{\parallel}^{3-2n},
\end{equation}
here $n$ denotes the index of the initial distribution function of
the resonant electrons ($\textit{f}_{\parallel_{0}}\propto
p_{\parallel}^{-n}$). From Eqs. (10) and (13) it follows that
$\textit{f}_{\parallel}(p_{\parallel})\propto p_{\parallel}^{n-2}$.
As the emitting particles in our case are the primary beam
electrons, nothing can be told about their initial distribution. We
only know the scenario of creation of the primary beam electrons
\cite{go69,besk} which are extracted from the pulsar's surface via
the electric field induced due to star rotation. To our knowledge
there is no convincing theory that would predict the initial form of
the distribution function of the beam electrons, which must be
drastically dependent on the neutron star's surface properties and
temperature.

The frequency of the original waves, excited during the cyclotron
resonance can be estimated from Eq. (2) as follows
\begin{equation}\label{42}
    \nu\approx2\pi\frac{\omega_{B}}{\delta\gamma_{b}}\sim10^{14}Hz.
\end{equation}
As we can see the frequency of the cyclotron modes comes in the same
domain as the measured optical emission of XDINSs with the certified
optical counterparts. Their spectra mostly follow the the
Rayleigh-Jeans tail $F_{\nu}\propto \nu^{2}$. On the other hand, the
spectral distribution of the cyclotron modes is given by expression
(13) and combining this equation with Eq. (14) we find
\begin{equation}\label{27}
     F_{\nu}\propto|E_{k}|^{2}\propto \nu^{2n-3}.
\end{equation}
From Eq. (15) it follows that when $n=5/2$ the spectral distribution
of the cyclotron modes is coincident with the Rayleigh-Jeans
function. And for the initial distribution of the beam electrons, as
for their final distribution we find
$\textit{f}_{\parallel_{0}}\propto p_{\parallel}^{-5/2}$ and
$\textit{f}_{\parallel}\propto p_{\parallel}^{1/2}$. We have used
this distribution for the beam electrons to define the theoretical
X-ray spectrum of RX J1856.5-3754, which well fitted the measured
one \cite{new}. Although, for RXJ0420 it is not yet confirmed the
detection of the optical counterpart, based on the similarity of
XDINSs we assume that the initial distribution function of the beam
electrons must be the same for all XDINSs. Thus, in place of
integral (8) as for two other XDINSs investigated in previous works
\cite{ch11,new} we get:
\begin{equation}\label{31}
    F_{\epsilon}\propto\epsilon^{0.3}
   \cdot e^{-\left(\epsilon/\epsilon_{m}\right)^{b}}.
\end{equation}
To find the values of the parameters $b$ and $\epsilon_{m}$, one
should perform a spectral analysis by fitting the model spectrum
absorbed by cold interstellar matter with the observed X-ray
spectrum of RXJ0420.

\section*{\sc spectral analysis}

To obtain the X-ray spectra of RXJ0420 with the highest statistical
quality we used the EPIC-pn data collected from three XMM-Newton
observations between December 2002 and January 2003. The archival
XMM-Newton data were processed with the Science Analysis Software
(SAS) version 11.0. The X-ray spectra were grouped in spectral bins
containing at least 30 photons. Subsequent spectral analysis was
performed with XSPEC V12.7.0. The three EPIC-pn spectra of RXJ0420
were fit simultaneously and spectral analysis was limited to
energies between $0.15$ and $1.0$ keV.  Along with the plasma
emission model proposed in the present work we also examined the
black-body model, to compare the fit results.

\subsection*{\sc the black-body model}

For the spectral analysis of EPIC-pn data for RXJ0420 first we used
an absorbed black-body model. We allowed only the black-body
normalization to vary between the spectra of the individual
observations and fitted the temperature and amount of interstellar
matter as common parameters. We found that the black-body provided a
reasonable fit, but with $N_{H}\sim10^{18}cm^{-2}$, that is not
perfect. Thus, we set a lower limit for this parameter, which did
not change drastically the fit quality. The resulting
$\chi^{2}=1.47$ for 97 degrees of freedom and for the column density
we got $N_{H}=(1.01\pm0.19)\times10^{20}cm^{-2}$. The best fit
black-body temperature $kT_{bb}^{\infty}=43$eV appears to be the
lowest value derived from any of the known XDINSs (see parameters in
Table 1).

\subsection*{\sc the synchrotron emission model}

The plasma emission model proposed in the present paper was recently
applied to explain the X-ray spectra of RX J1856.5-3754 and RBS1774,
revealing good fit quality in both cases \cite{ch11,new}. We
performed fitting of the model spectrum (Eq.(16)) absorbed by cold
interstellar matter with the EPIC-pn spectra of RXJ0420.  The best
fit results are $b=1.27$ and $\epsilon_{m}=0.1$keV, corresponding
$\chi^{2}=1.51$ for  96 degrees of freedom. The column density
$N_{H}$, $\epsilon_{m}$ and $b$ were treated as free parameters
common to all three spectra. Only normalization was allowed to vary
freely for different spectra independently, as in previous case when
applying the black-body model. The fit results are listed in Table
1.

\begin{table*}
\label{table} \caption{Model-Fit spectral parameters of RXJ0420.0-5022 for combined fits to the EPIC-pn spectra obtained from
individual observations in the energy interval $0.15-1.0$ keV.}
\centering
\begin{tabular}{llrrrrlr}
\hline \\
Model    & $N_{H}$ & $\epsilon_{m}$ & $b$ & $kT_{bb}^{\infty}$ & $E_{edge}$ & $\tau_{edge}$ & $\chi^{2}$(dof)    \\
& $(10^{20}cm^{-2})$ & (keV) && (eV)  & (keV) &\\
\hline
\hline    \\                    
plasma & $1.00^{+0.28}_{-0.28}$ & $0.10\pm0.04$& $1.27\pm0.31$ &&&& $1.51(96)$\\ \\
plasma*edge & $1.00^{+0.37}_{-0.37}$ & $0.10\pm0.05$& $1.27\pm0.42$ & & $0.30^{+0.01}_{-0.01}$ & $0.68^{+0.30}_{-0.30}$  & $1.10(94)$\\ \\
bbody & $1.01^{+0.19}_{-0.19}$ &&  & $43.3\pm1.2$ &  &  &    $1.47(97)$\\
\\
bbody*edge & $1.01^{+0.33}_{-0.33}$&&  &  $46.3\pm1.8$ & $0.31^{+0.02}_{-0.02}$ & $0.69^{+0.21}_{-0.21}$  & $1.10(95)$\\
\hline
\end{tabular}
\end{table*}

\begin{figure*}[!h]
\centering
\epsfig{file=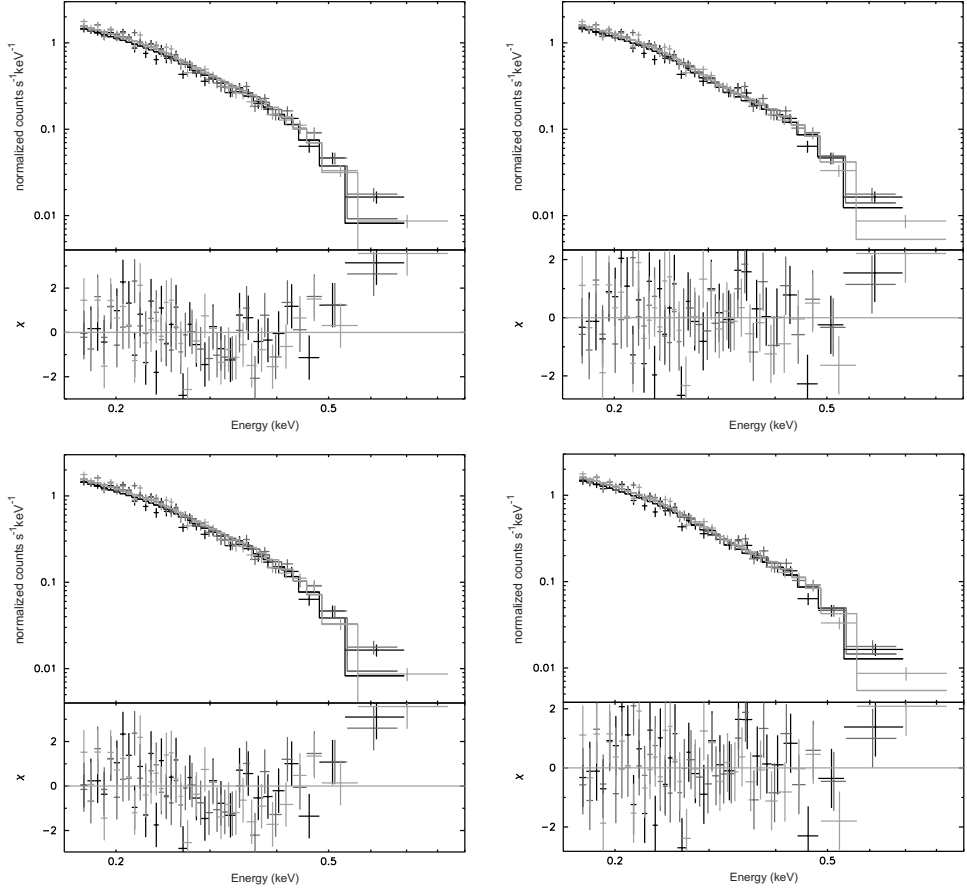,width=0.7\linewidth}
\caption{Combined EPIC-pn spectra of RX J0420.0-5022. Top left: The absorbed black-body model fit. Top right: The black-body model fit including an absorption edge at $\sim0.3$keV. Bottom left: The synchrotron emission model fit  absorbed by cold interstellar matter. Bottom right: Model fit including absorption edge at 0.3keV.}
\label{1}
\end{figure*}

\section*{\sc discussion}

The spectral analysis of the pulse phase-averaged X-ray spectra of RXJ0420 shows that the fits quality in case of both models
(black-body and plasma emission models) is not very good. From inspection of Fig.1 one might consider the residuals around $0.3$keV
as an absorption feature. This feature, described as a broad ($\sigma=70$eV) Gaussian absorption line was noted and discussed by
\cite{hab04}. We add the absorption line at $\sim0.3$keV to both models considered in the present work and re-fit the data. The fit
quality improves in both cases, only if using an absorption edge model. The best-fitting energy of the edge is $E_{edge}\approx0.3$keV, and the optical depth  $\tau_{edge}\approx0.7$  (see Tab.1). The resulting value of $\chi^{2}$ reduces to $1.1$ for both models.

The nature of the spectral features discovered in X-ray spectra of XDINSs is not fully clarified as yet. According to the thermal
emission scenarios the most likely interpretation of the absorption lines is that they appear due to proton cyclotron resonance. The
proton cyclotron absorption line at $\sim0.3$keV implies the magnetic field of $B_{cyc}=6.6\times10^{13}$G, which differs from
the value of the dipolar magnetic field inferred from the timing measurements $B_{dip}=1.0\times10^{13}$G \cite{hab07}.

We suppose that existence of the absorption feature in X-ray spectra of RXJ0420 might be caused by wave damping at photon energies
$\sim0.3$keV, which takes place near the light cylinder. During the farther motion in the pulsar magnetosphere, the X-ray emission might come in the cyclotron damping range. If we assume that damping happens on the left slope of the distribution function of primary beam electrons (see Fig. 1 from \cite{arons}), then the photon energy of damped waves will be $\epsilon_{0}=(h/2
\pi)2\omega_{B}/\gamma_{b}\psi^{2}\simeq 0.3$keV \cite{ch11}. Taking into account the shape of the distribution function of beam
electrons, we interpret the large residuals around $\sim0.3$keV as an absorption edge.

Despite the fact that fit quality considerably improves when
includin the absorption edge at $0.3$keV for both emission models,
the physical reality of this feature is still uncertain. It might be
caused by calibration uncertainties. A feature of possible similar
nature was detected in EPIC-pn spectra of the much brighter
prototypical object RXJ1856.4-3754 and classified as remaining
calibration problem by \cite{hab07}. Thus, we agree that more data
are necessary to finally prove or disprove the existence of this
feature.

According to the fit results (see Tab.1) the spectral analysis of
the measured EPIC-pn X-ray spectra of RXJ0420 does not seem to be
enough to distinguish between the black-body and the plasma emission
models. Differently from the thermal radiation models, which appear
to be somehow artificial, our scenario is based on a self-consistent
theory.  According to works of \cite{sturrock71} and \cite{tadem}
due to the cascade processes of a pair creation the pulsar's
magnetosphere is filled by electron-positron plasma with  the
anisotropic one-dimensional distribution function. The beam
particles undergo drifting perpendicularly to the magnetic field due
to the curvature, of the field lines. Both of these factors (the
one-dimensionality of the distribution function and the drift of
particles) might cause the generation of eigen modes in the
electron-positron plasma if the condition of cyclotron resonance is
fulfilled. The generated waves interact with the resonant electrons
via the QLD, which leads to the diffusion of particles as along as
across the magnetic field lines, and inevitably causes creation of
pitch angles by resonant particles. Therefore, the resonant
electrons start to radiate in the synchrotron regime.

The estimations show that for the beam electrons with the
average Lorentz-factor $\gamma_{b}\sim10^{7}$, the synchrotron radiation
comes in the same domain as the measured X-ray spectrum of RXJ0420.
Differently from the standard theory of the synchrotron emission
\cite{ginz81}, which only provides the power-law spectrum,
our approach gives the possibility to obtain different spectral energy
distributions. In the standard theory of the synchrotron emission,
it is supposed that the observed radiation is collected from a large
spacial region in various parts of which, the magnetic field is
oriented randomly. Thus, it is supposed that along the line of sight
of an observer the magnetic field directions  are chaotic.
In our case the emission comes from the region of the pulsar magnetosphere
where the magnetic field lines are practically straight and parallel to each other. And differently from standard approach, we take into account the mechanism of creation of the pitch angles, which inevitably restricts their possible values.

\section*{\sc acknowledgement}
\indent \indent The authors are grateful to Prof. George Machabeli
for valuable discussions. The HEASARC online data archive at
NASA/GSFC has been used extensively in this research.

\end{document}